\documentclass[conference]
{IEEEtran}
\IEEEoverridecommandlockouts
\usepackage{cite}
\usepackage{multicol}
\usepackage{amsmath,amssymb,amsfonts}
\usepackage{algorithmic}
\usepackage{balance}
\usepackage{graphicx}
\usepackage{subfig}
\usepackage{hyperref}
\usepackage{textcomp}
\usepackage{xcolor}
\usepackage{booktabs}
\usepackage{multirow}
\hypersetup{
    colorlinks=true,
    linkcolor=blue,
    citecolor=blue,
    urlcolor=blue
}
\usepackage[linesnumbered,ruled,vlined]{algorithm2e}
\def\BibTeX{{\rm B\kern-.05em{\sc i\kern-.025em b}\kern-.08em
    T\kern-.1667em\lower.7ex\hbox{E}\kern-.125emX}}
\usepackage{fancyhdr}

\begin{document}
\title{FedClust: Optimizing Federated Learning on Non-IID Data through Weight-Driven Client Clustering \\
}
\author{
    \IEEEauthorblockN{Md Sirajul Islam\textsuperscript{1}, Simin Javaherian\textsuperscript{1}, Fei Xu\textsuperscript{2}, Xu Yuan\textsuperscript{3}, Li Chen\textsuperscript{1}, and Nian-Feng Tzeng\textsuperscript{1}}
    \IEEEauthorblockA{\textsuperscript{1}School of Computing and Informatics, University of Louisiana at Lafayette, USA\\
    \textsuperscript{2}School of Computer Science and Technology, East China Normal University, China\\
    \textsuperscript{3}Department of Computer and Information Sciences, University of Delaware, USA\\
    Emails: \{md-sirajul.islam1, simin.javaherian1, li.chen, tzeng\}@louisiana.edu, fxu@cs.ecnu.edu.cn, xyuan@udel.edu}
}
\maketitle
\thispagestyle{fancy}
\begin{abstract}
Federated learning (FL) is an emerging distributed machine learning paradigm enabling collaborative model training on decentralized devices without exposing their local data. A key challenge in FL is the uneven data distribution across client devices, violating the well-known assumption of independent-and-identically-distributed (IID) training samples in conventional machine learning. Clustered federated learning (CFL) addresses this challenge by grouping clients based on the similarity of their data distributions. However, existing CFL approaches require a large number of communication rounds for stable cluster formation and rely on a predefined number of clusters, thus limiting their flexibility and adaptability. This paper proposes {\em FedClust}, a novel CFL approach leveraging correlations between local model weights and client data distributions. {\em FedClust} groups clients into clusters in a one-shot manner using strategically selected partial model weights and dynamically accommodates newcomers in real-time. Experimental results demonstrate {\em FedClust} outperforms baseline approaches in terms of accuracy and communication costs. 

\end{abstract}


\section{Introduction}
Federated learning (FL) has become a promising solution for analyzing and processing the massive data generated by various edge devices. Traditional machine learning approaches fall short in handling this exponential increase in data, as they require transmitting large volumes of user data to centralized cloud servers. This incurs prohibitive communication costs and raises privacy concerns. FL enables a set of participating devices to collaboratively train a globally shared model under the coordination of a central server, without exposing their local data. Due to its superior privacy-preservation implications, FL has been widely adopted in diverse areas.

However, deploying FL often involves a significant number of devices that generate heterogeneous data due to varying usage patterns of users. The presence of heterogeneous data across client devices breaks the conventional assumption of independent-and-identically-distributed (IID) training data, raising the new challenge of non-IID data distribution in the FL paradigm.
Such a data heterogeneity issue not only increases the overall communication cost but also degrades global model performance \cite{li2019convergence, zhao2018federated}, increasingly drawing research attention to mitigate the adverse impact of non-IID data on FL \cite{ghosh2020efficient, ouyang2021clusterfl, li2020}.
Several FL approaches, including personalization \cite{ fallah2020personalized, li2021ditto, luo2022pgfed}, clustering \cite{ghosh2020efficient, ouyang2021clusterfl, vahidian2022efficient}, and device selection \cite{cho2020client, lai2021oort, javaherian2024fedfair} have been proposed in the literature to alleviate the impact of non-IID data. Despite the significant improvements CFL-based approaches have shown over FedAvg \cite{mcmahan2017communication} in handling non-IID data, they still lack efficiency due to limitations in clustering strategies. We identify key limitations of existing CFL methods, including the difficulty in pre-determining cluster counts \cite{ghosh2020efficient, ouyang2021clusterfl}, the need for larger communication rounds to form stable clusters \cite{sattler2020clustered}, and the necessity of utilizing all model weights for clustering \cite{ghosh2020efficient, sattler2020clustered}.  

\begin{figure*}
\centering
\subfloat[Layer 1 (CL)]
{\includegraphics[height=3.5cm, width=0.23\textwidth]{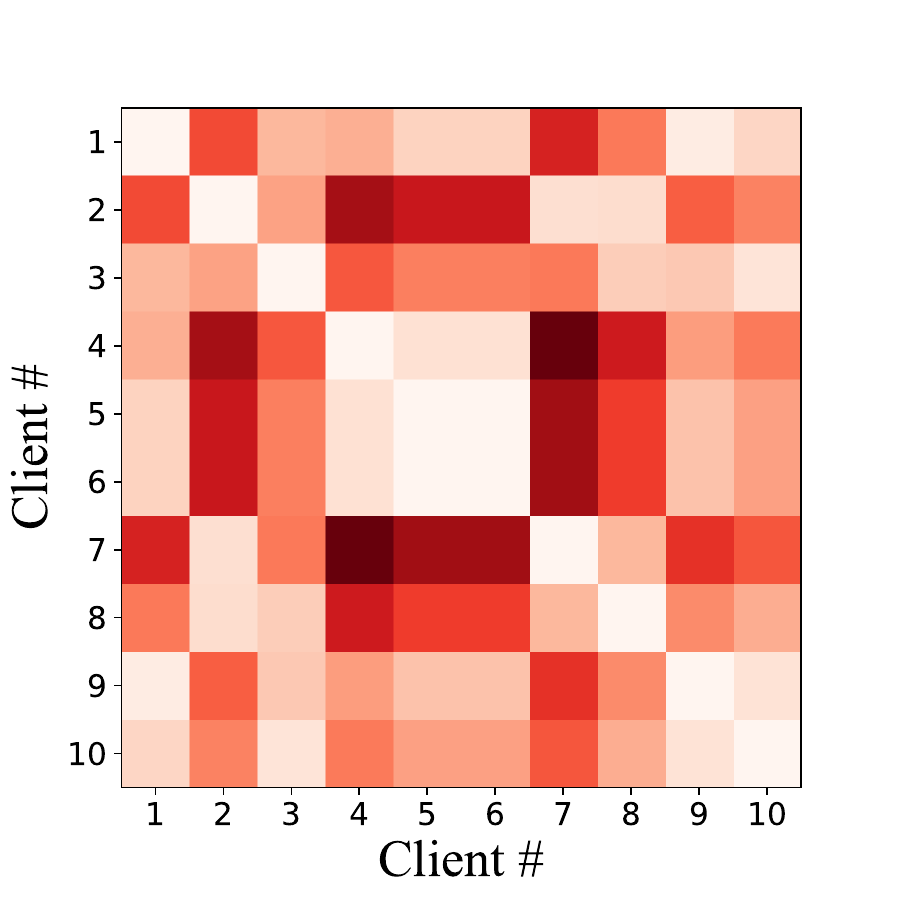}\label{fig:1}}
\subfloat[Layer 7 (CL)]
{\includegraphics[height=3.5cm, width=0.23\textwidth]{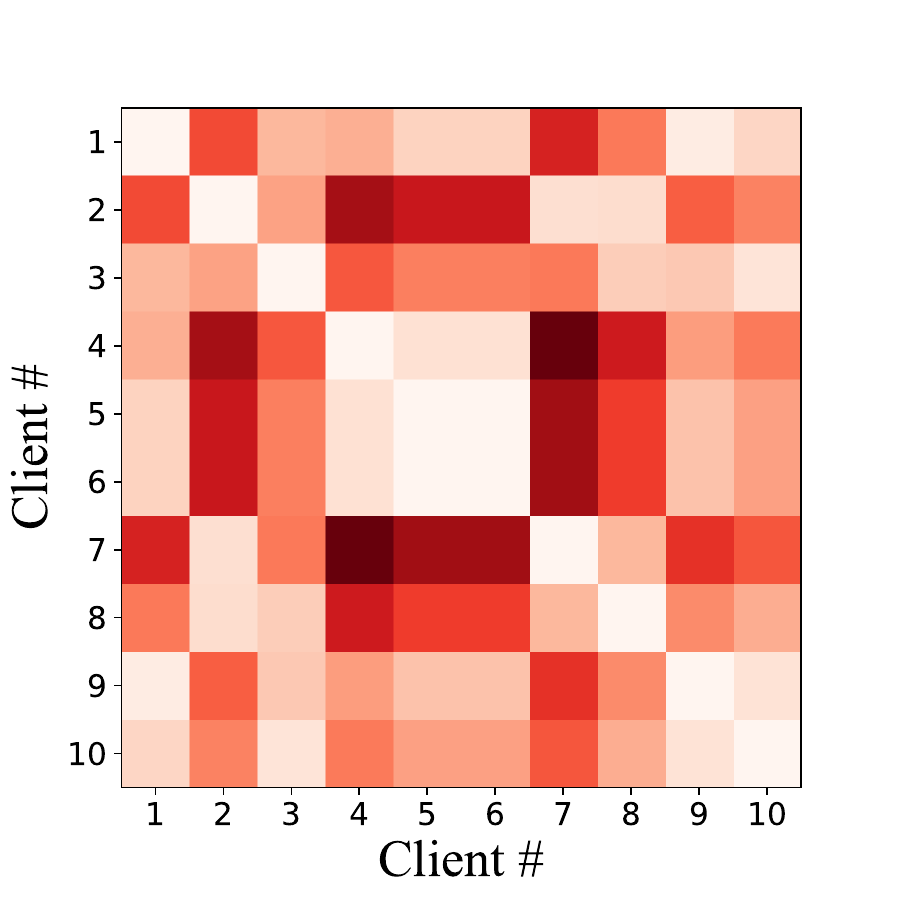}\label{fig:2}}
\subfloat[Layer 14 (FL)]
{\includegraphics[height=3.5cm, width=0.23\textwidth]{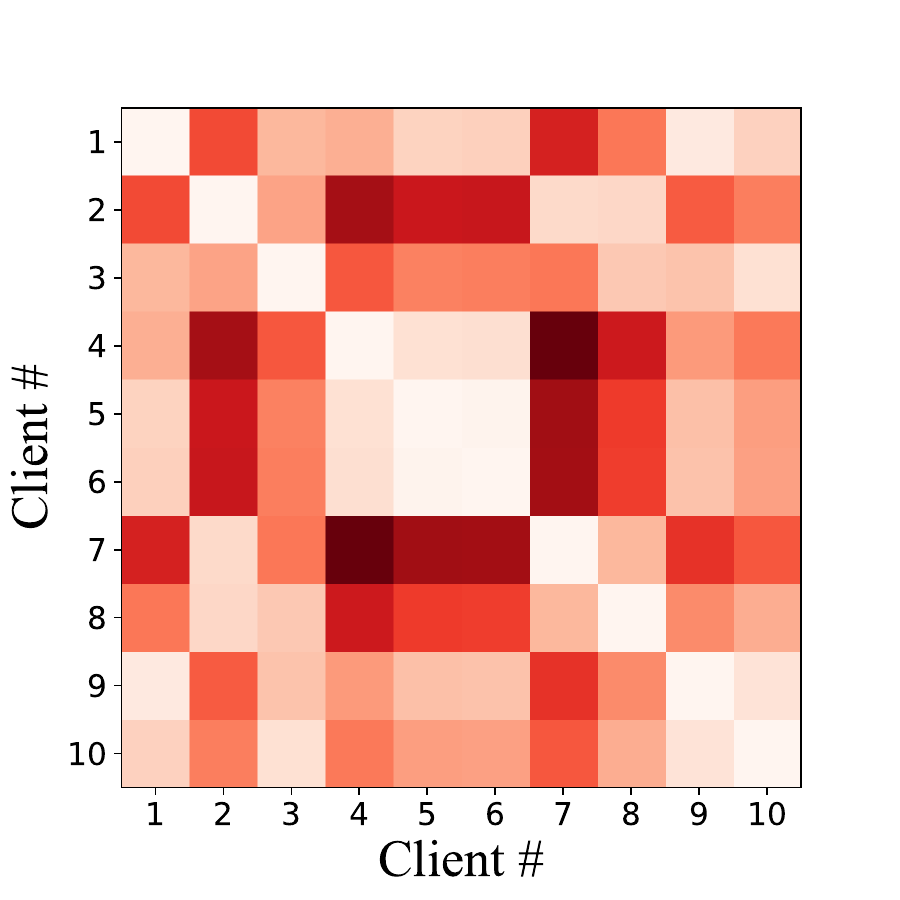}\label{fig:3}}
\subfloat[Layer 16 (FL)]
{\includegraphics[height=3.5cm, width=0.23\textwidth]{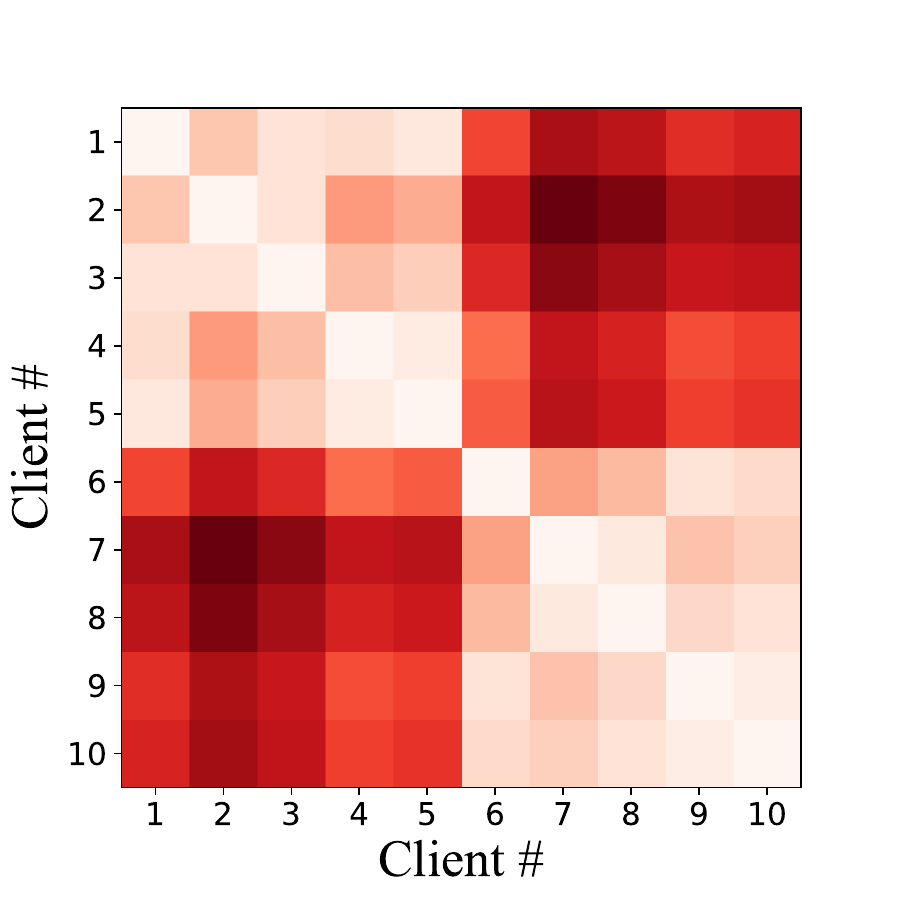}\label{fig:4}}
\caption{Illustration of the distance matrices calculated using different layer weights (CL: convolutional layer, FL: fully connected layer). Lighter color indicates smaller distances, i.e., the two models are more similar. }
\label{figure}
\end{figure*}

To address the aforementioned limitations, we propose a novel clustered federated learning method, named {\em FedClust}, which leverages our insight into the implicit relationship between the local model weights and the underlying data distribution on a client device. 

\section{Background and Motivation}

Federated learning (FL) is a privacy-preserving framework enabling distributed clients to collaboratively train a shared global model without sharing their local data. In FL, clients receive initial global model parameters from a central server, train it for a few local iterations using their local data, and send model updates back to the server for aggregation. After aggregation, the server sends it to clients for further training. The above process continues until achieving a certain level of model accuracy or a pre-specified number of communication rounds. The server has no prior knowledge about the data distribution across devices as it cannot access the raw data stored in clients.
In particular, the goal of {\em FedAvg} is typically to minimize the following objective function:
\begin{equation}
\min_{\theta} F(\theta) \overset{\scriptscriptstyle\Delta}{=} \sum_{i=1}^{m}\frac{n_i}{N}F_i(\theta)
\end{equation}
Here, $m$ is the set of participating clients and client $i$ has local dataset $\mathcal{D}_i$, where $n_i =|\mathcal{D}_i|$ and $N = \sum_{i=1}^{m} n_i$. The local objective functions of clients can be defined as the empirical loss over their local data $\mathcal{D}_i$, i.e., $F_i(\theta)=\frac{1}{n_i}\sum_{j_i=1}^{n_i}f_{j_i} (\theta; x_{j_i},y_{j_i})$, where $n_i$ is the number of each client local samples. 
\begin{figure}[b]
\centering
\includegraphics[height=5cm, width=0.5\textwidth]{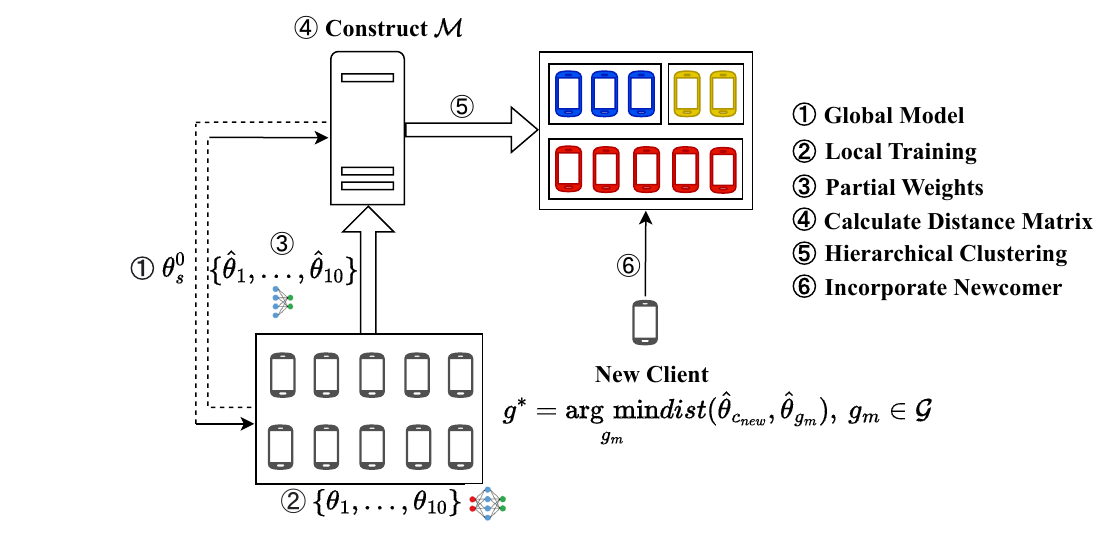}
\caption{An overview of {\em FedClust}. }
\label{figure1}
\end{figure}
We experimentally investigate the impact of model weights from various layers on the underlying data distribution in FL by performing a multi-class image classification task on the CIFAR-10 \cite{krizhevsky2009learning} dataset with VGG16 \cite{simonyan2014very}. To simulate non-IID data among clients, we assume 10 different clients and group them into two clusters based on their local label categories, {\em e.g.}, $\mathcal{G}_1 = \{1, 2, \ldots, 5\}$ and $\mathcal{G}_2 = \{6, 7, \ldots, 10\}.$ From Fig. 1, we observe that the final layer weights implicitly represent the underlying data distribution of clients. The clustering structure of the clients is clearly observed in Fig. 1(d). Based on the above experiment and previous studies \cite{long2018transferable, yosinski2014transferable}, we can conclude that the final layer or the layer with the classifier function reflects the model difference caused by non-IID data. In addition, clients with similar data distributions tend to train the model in a similar manner, resulting in closer distances among final layer weights.

\section{The framework of FedClust}
We present the framework of the proposed FedClust as follows. As shown in Fig. 2, the framework of FedClust can be described as follows.
\begin{itemize}
\item First, the server broadcasts the initial global model parameters to all available clients. 

\item The participating clients locally train the model based on their data, and upload the updated final layer weights to the server to represent their underlying data distribution.
\item The server then computes the proximity matrix between models based on the Euclidean distance among final layer weights uploaded by each client.
\item Finally, the server employs agglomerative hierarchical clustering (HC) \cite{day1984efficient} on the proximity matrix $\mathcal{M}$ to group clients with similar data distribution into the same cluster. 

\item The above clustering process is done in one communication round. From the next round, the workflow of {\em FedClust} is similar to {\em FedAvg} \cite{mcmahan2017communication}.
\end{itemize}
\begin{table}[t]
\caption{Test accuracy comparisons of different approaches over different datasets for Non-IID Dir (0.1).}
\centering
\scriptsize
\resizebox{\columnwidth}{!}{\begin{tabular}{lllll}
\toprule
\text{Method} & \text{CIFAR-10} & \text{FMNIST} &\text{SVHN}\\
\midrule
FedAvg&38.25 ± 2.98 & 81.93 ± 0.64 &  61.26 ± 0.95 \\
FedProx&51.60 ± 1.40 & 74.53 ± 2.16 & 79.64 ± 0.80 \\
CFL&41.50 ± 0.35 & 74.01 ± 1.19 &  61.96 ± 1.58  \\
IFCA&50.51 ± 0.61 & 84.57 ± 0.41 & 74.57 ± 0.40 \\
PACFL&51.02 ± 0.24 & 85.30 ± 0.28 & 76.35 ± 0.46 \\
\textbf {FedClust}&\textbf{60.25 ± 0.58} & \textbf{95.51 ± 0.17} & \textbf{78.23 ± 0.30} \\
\bottomrule
\end{tabular}}
\end{table}
\section{Performance Evaluation}
To evaluate the performance of {\em FedClust}, we consider LeNet-5 \cite{lecun1989backpropagation} model on CIFAR-10 \cite{krizhevsky2009learning}, Fashion MNIST (FMNIST) \cite{xiao2017fashion}, and SVHN \cite{netzer2011reading} datasets. We simulate non-IID scenarios using Non-IID Dir (0.1) data heterogeneity settings as in \cite{li2022federated}. We consider following baseline approaches: {\em FedAvg} \cite{mcmahan2017communication}, {\em FedProx} \cite{li20201federated}, {\em IFCA} \cite{ghosh2020efficient}, {\em PACFL} \cite{vahidian2022efficient} and {\em CFL} \cite{sattler2020clustered}. Results in Table I demonstrate the effectiveness of {\em FedClust} over these baselines. Exploring the performance of {\em FedClust} across various data heterogeneity scenarios with different models and datasets has been left as part of our future work.
\section{Acknowledgement}
The research is supported in part by the NSF under grants OIA-2019511, OIA-2327452, 2348452, and 2315613, in part by the Louisiana BoR under LEQSF(2019-22)-RD-A-21, in part by the NSFC under 62372184, and the Sci. and Tech. Commission of Shanghai Municipality under 22DZ2229004.
\balance

\bibliographystyle{IEEEtran}
\bibliography{simple-base}
\end{document}